\documentclass[11pt,twoside]{article}
\usepackage{asp2014}

\aspSuppressVolSlug
\resetcounters

\begin{document}

\title{Field tests for the ESPRESSO data analysis software}
\author{Guido~Cupani,$^1$ Valentina~D'Odorico,$^1$ Stefano~Cristiani,$^1$ Jonay~I.~Gonz\'alez~Hern\'andez,$^2$ Christophe~Lovis,$^3$ S\'ergio~Sousa,$^4$ Paolo~Di~Marcantonio,$^1$ and Denis~M\'egevand$^3$
\affil{$^1$INAF--OATs, Via Tiepolo 11, 34143 Trieste, Italy}
\affil{$^2$IAC, V\'ia L\'actea, 38205 La Laguna, Tenerife, Spain}
\affil{$^3$Universit\'e de Gen\`eve, 51 Chemin des Maillettes, 1290 Versoix, Switzerland}
\affil{$^4$Instituto de Astrof\'isica e Ci\^encias do Espa\c{c}o, Universidade do Porto, CAUP, Rua das Estrelas, 4150-762 Porto, Portugal}}

\paperauthor{Guido~Cupani}{cupani@oats.inaf.it}{}{INAF}{OATs}{Trieste}{}{34143}{Italy}
\paperauthor{Valentina~D'Odorico}{dodorico@oats.inaf.it}{}{INAF}{OATs}{Trieste}{}{34143}{Italy}
\paperauthor{Stefano~Cristiani}{cristiani@oats.inaf.it}{}{INAF}{OATs}{Trieste}{}{34143}{Italy}
\paperauthor{Jonay~I.~Gonz\'alez~Hern\'andez}{jonay@iac.es}{}{IAC}{}{La Laguna}{Tenerife}{38205}{Spain}
\paperauthor{Christophe~Lovis}{Christophe.Lovis@unige.ch}{}{Universit\'e de Gen\`eve}{}{Versoix}{}{1290}{Switzerland}
\paperauthor{S\'ergio~Sousa}{sousasag@astro.up.pt}{}{Universidade do Porto}{Instituto de Astrof\'isica e Ci\^encias do Espa\c{c}o}{Porto}{}{4150-762}{Portugal}
\paperauthor{Paolo~Di~Marcantonio}{dimarcan@oats.inaf.it}{}{INAF}{OATs}{Trieste}{}{34143}{Italy}
\paperauthor{Denis~M\'egevand}{denis.megevand@unige.ch}{}{Universit\'e de Gen\`eve}{}{Versoix}{}{1290}{Switzerland}

\begin{abstract}
The data analysis software (DAS) for VLT ESPRESSO is aimed to set a new benchmark in the treatment of spectroscopic data towards the extremely-large-telescope era, providing carefully designed, fully interactive recipes to take care of complex analysis operations (e.g. radial velocity estimation in stellar spectra, interpretation of the absorption features in quasar spectra). A few months away from the instrument's first light, the DAS is now mature for science validation, with most algorithms already implemented and operational. In this paper, I will showcase the DAS features which are currently employed on high-resolution HARPS and UVES spectra to assess the scientific reliability of the recipes and their range of application. I will give a glimpse on the science that will be possible when ESPRESSO data become available, with a particular focus on the novel approach that has been adopted to simultaneously fit the emission continuum and the absorption lines in the Lyman-alpha forest of quasar spectra.
\end{abstract}

\section*{ESPRESSO in a nutshell}
ESPRESSO \citep{2013Msngr.153....6P} is an ultra-stable, high-resolution spectrograph ($R\sim 55,000$ to $200,000$) for the coud\'{e} combined focus of the Very Large Telescope (VLT) of the European Southern Observatory (ESO). Its driving scientific objectives are (i) the search for Earth-like exoplanets and (ii) the exploration of new physics beyond the standard model, through a measure of the possible variation of the dimensionless constants $\alpha$ (fine-structure constant) and $\mu$ (proton-to-electron mass ratio). The latter science case depends on the accurate analysis of the absorption features produced by the inter-galactic and circum-galactic medium on the spectrum of background bright sources such as quasars (QSOs). The same analysis provides a valuable insight into the physical and chemical state of the baryonic matter from the reionization epoch onwards and on its interplay with galaxies.

Since the inception of its development, ESPRESSO has been conceived as a ``science machine'' able to produce scientific results within minutes from the end of observations. To this aim the instrument is equipped with dedicated software tools to handle both the data reduction and the data analysis, the latter covering both stellar and QSO spectral analysis \citep{2012SPIE.8448E..1OD,2014SPIE.9149E..1QD}. The ESPRESSO Data Analysis Software (DAS) has been introduced in a series of previous articles \citep{2015MmSAI..86..502C,2015ASPC..495..289C,2016SPIE_cupani_1}. In this article we discuss the status of its development at the instrument integration stage and present some results of the first science assessment on test data, a few months before ESPRESSO is commissioned and sees its first light. We will focus in particular on the QSO spectral analysis, as it embraces some of the most interesting feature of the DAS, both in its algorithms and interface.

\section*{The DAS concept}
The ESPRESSO DAS is meant to set a benchmark in the treatment of spectroscopic data towards the ELT era, providing carefully designed, fully interactive recipes to take care of complex analysis operations. Those are (i) for stellar spectra: computation of the radial velocity, the stellar activity indexes, the equivalent width of absorption lines, and the stellar parameters (effective temperature, [Fe$/$H]); continuum fitting and re-computation of the radial velocity by comparison with synthetic spectra; (ii) for QSO spectra: detection of the absorption lines; determination of the emission continuum level; identification and fitting of the absorption systems. 

Together with the DRS, the DAS enforce a ``pixel conservation'' paradigm, in which the information collected by the individual pixels of the detector is preserved throughout the reduction cascade \citep{2016SPIE_cupani_2}. Whenever the information from different pixels is merged, potentially disrupting the flux statistics (such as in re-binning and co-addition of multiple exposures), the software propagates the information both in merged and non-merged form, to allow for a correct assessment of theoretical models (such as Voigt profiles for absorption lines) with standard best-fit techniques.

The bulk of the DAS code is written in ANSI-C. Most of the code is developed using the ESO Common Pipeline Library \citep{2004SPIE.5493..444M} which provides low-level tool for data handling; a library of higher-level functions has been designed to address individual tasks (e.g. spectral rebinning, curve smoothing, line fitting) which are organized into self-standing modules (``recipes''). The cascade of recipes is typically run within the ESO Reflex environment \citep{2013A&A...559A..96F} which provides an intuitive workflow interface complemented with Python scripts to visualize the results and adjust the parameters.

\articlefigure[width=\textwidth]{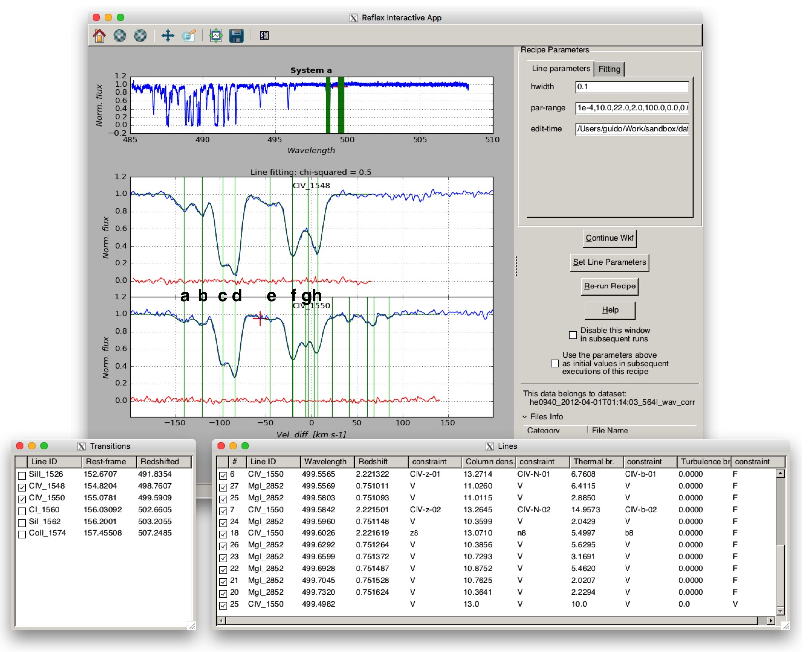}{line_fit}{Reflex interactive windows for line fitting, with a fitted C \textsc{iv} doublet in the spectrum of QSO HE 0940$-$1050. The bottom left panel shows the list of detected transitions (displayed transitions checked); the bottom right panel shows the list of available components (fitted components checked). The user may add components by clicking in the spectrum plot; when this is done, the panels are consistently updated at run-time. The labels on the paired components are given for reference to Table \ref{fitlyman}.}

\begin{table}[!ht]\caption{Line parameters fitted by the DAS on the paired components of the C \textsc{iv} doublet shown in Fig.~\ref{line_fit}, with the $1\sigma$ uncertainty of the last significant digit(s) in parentheses. The component IDs correspond to the labels in figure. $z$: line redshift; $N$: Voigt-profile column density; $b$: Voigt-profile thermal broadening (turbulence broadening has been neglected). Next to each parameter, we listed the difference (when measurable) between the values fitted by the DAS and those fitted by the ESO MIDAS FITLYMAN package. In all cases, this difference is well below the fit uncertainty.}\label{fitlyman}
\begin{center}
\begin{tabular}{cr@{\,}l@{\;\;}cr@{\,}l@{\;\;}cr@{\,}lc}
\hline
Comp. & \multicolumn{2}{c}{$z$} & $\Delta z$ & \multicolumn{2}{c}{$\log{N}$} & $\Delta\log{N}$ & \multicolumn{2}{c}{$b$} & $\Delta b$\\ 
& & & [$10^{-6}$] & & & [$10^{-2}$] & \multicolumn{2}{c}{[km s$^{-1}$]} & [$10^{-1}$ km s$^{-1}$]\\
\hline
a & $2.220040$ & $(6)$  & $-1$ & $12.65$ & $(3)$ &      & $ 9.5$ & $(9)$   & $-1$\\
b & $2.220253$ & $(4)$  & $-1$ & $12.66$ & $(3)$ &      & $ 6.7$ & $(6)$   & $-1$\\
c & $2.220501$ & $(2)$  &      & $13.48$ & $(2)$ &      & $ 6.8$ & $(3)$   & $-2$\\
d & $2.220645$ & $(2)$  &      & $13.58$ & $(1)$ &      & $ 5.8$ & $(2)$   & $-1$\\
e & $2.221055$ & $(11)$ & $-2$ & $12.63$ & $(4)$ & $-1$ & $14.0$ & $(1.6)$ & $-5$\\
f & $2.221322$ & $(3)$  &      & $13.27$ & $(7)$ &      & $ 6.8$ & $(5)$   & $+1$\\
g & $2.221501$ & $(36)$ & $-3$ & $13.26$ & $(5)$ & $-1$ & $15.0$ & $(2.0)$ & $+1$\\
h & $2.221619$ & $(3)$  & $+1$ & $13.07$ & $(7)$ & $+2$ & $ 5.5$ & $(6)$   & $-1$\\
\hline
\end{tabular}
\end{center}
\end{table}

\section*{Validating the analysis of QSO spectra}

As of November 2016, eight internal releases of the DAS has been issued for verification by ESO. All but one recipes have been coded and three out of four Reflex workflows (two for the stellar branch, one for the QSO branch) are already in operation. As the integration of the instrument progresses, the code is being validated both on reduced high-resolution spectra from VLT UVES and TNG HARPS-North (for scientific purpose) and on the first ESPRESSO test data processed by the DRS (to check the DRS/DAS interface). The first public release is foreseen for the instrument commissioning (2017).

The QSO branch of the DAS includes new algorithms to automatically fit the continuum emission and  the absorption features produced by the intervening structures. By nature, such analysis can only be performed iteratively: the continuum level is determined by fitting and removing all the absorption lines, while the lines are fitted with respect to a previously determined continuum. In practice, the workflow operates as follows: (i) the continuum is first determined by making initial assumptions on the nature of lines (distribution of column densities in the forest of Lyman-$\alpha$ absorbers; guesses on other Voigt parameters) and then the spectrum is normalized; (ii) associated lines (corresponding to different atomic transitions at the same redshift) are selected to define absorption systems; (iii) absorption systems are then fitted with Voigt profiles, adjusting the number of components and the constraint among line parameters. The information from line fitting can finally be used to refine the continuum estimation. Both continuum and line fitting are validated (by a $\chi^2$ test) on the non-rebinned spectra, according to the pixel conservation paradigm, to allow a correct modelling of the flux variance. 

Multiple test have been conducted on observations of QSO HE 0940$-$1050 (observed with UVES). The automatic continuum estimation in the Lyman-$\alpha$ forest (which includes an estimation of the residual optical depth not accounted for by fitted lines, modelled from the distribution of the neutral hydrogen column density) is consistent with the results of visual estimation, providing in addition a $\chi^2$ goodness-of-fit assessment. Within the Reflex environment, we developed a user-friendly interface to line fitting which allows to interactively select the transitions and set up constraints among the Voigt parameters (Fig.~\ref{line_fit}). Comparison with other packages for Voigt-profile fitting (such as the ESO MIDAS FITLYMAN package, \citealt{1995Msngr..80...37F}; see table \ref{fitlyman}) shows perfect consistence. More tests are currently ongoing, taking advantage of a larger UVES test data set.

\bibliography{P1-13} 

\begin{thebibliography}{}
\expandafter\ifx\csname natexlab\endcsname\relax\def\natexlab#1{#1}\fi
\expandafter\ifx\csname url\endcsname\relax
  \def\url#1{\texttt{#1}}\fi
\expandafter\ifx\csname urlprefix\endcsname\relax\def\urlprefix{URL }\fi
\providecommand{\eprint}[2][]{\url{#2}}

\bibitem[{{Cupani} et~al.(2015{\natexlab{a}})}]{2015ASPC..495..289C}
{Cupani}, G., et~al. 2015{\natexlab{a}}, in ADASS XXIV, edited by A.~R.
  {Taylor}, \& E.~{Rosolowsky}, vol. 495 of ASP Conf. Ser., 289

\bibitem[{{Cupani} et~al.(2015{\natexlab{b}})}]{2015MmSAI..86..502C}
--- 2015{\natexlab{b}}, Mem. Societ\`a Astronomica Italiana, 86, 502

\bibitem[{{Cupani} et~al.(2016{\natexlab{a}})}]{2016SPIE_cupani_1}
--- 2016{\natexlab{a}}, in Software and Cyberinfrastructure for Astronomy III,
  edited by G.~{Chiozzi}, \& J.~C. {Guzman}, vol. 9913 of Proc. SPIE, 99131T

\bibitem[{{Cupani} et~al.(2016{\natexlab{b}})}]{2016SPIE_cupani_2}
--- 2016{\natexlab{b}}, in Observatory Operations: Strategies, Processes, and
  Systems V, edited by A.~B. {Peck}, \& R.~L. {Seaman}, vol. 9910 of Proc.
  SPIE, 991023

\bibitem[{{Di Marcantonio} et~al.(2012)}]{2012SPIE.8448E..1OD}
{Di Marcantonio}, P., et~al. 2012, in Observatory Operations: Strategies,
  Processes, and Systems IV, edited by A.~B. {Peck}, R.~L. {Seaman}, \&
  F.~{Comeron}, vol. 8448 of Proc. SPIE, 84481O

\bibitem[{{Di Marcantonio} et~al.(2014)}]{2014SPIE.9149E..1QD}
--- 2014, in Observatory Operations: Strategies, Processes, and Systems V,
  edited by A.~B. {Peck}, C.~R. {Benn}, \& R.~L. {Seaman}, vol. 9149 of Proc.
  SPIE, 91491Q

\bibitem[{{Fontana} \& {Ballester}(1995)}]{1995Msngr..80...37F}
{Fontana}, A., \& {Ballester}, P. 1995, The Messenger, 80, 37

\bibitem[{{Freudling} et~al.(2013)}]{2013A&A...559A..96F}
{Freudling}, W., et~al. 2013, \aap, 559, A96

\bibitem[{{McKay} et~al.(2004)}]{2004SPIE.5493..444M}
{McKay}, D.~J., et~al. 2004, in Optimizing Scientific Return for Astronomy
  through Information Technologies, edited by P.~J. {Quinn}, \& A.~{Bridger},
  vol. 5493 of Proc. SPIE, 444

\bibitem[{{Pepe} et~al.(2013)}]{2013Msngr.153....6P}
{Pepe}, F., et~al. 2013, The Messenger, 153, 6

\end{thebibliography}

\end{document}